\documentstyle[amssymb,preprint,aps,prl,epsfig]{revtex}
\draft

\begin{document}
\title{dimerization in a half-filled one-dimensional extended Hubbard model}
\author{Y. Z. Zhang}
\address{Max-Planck-Institut f\"{u}r Physik komplexer Systeme, \\
N\"othnitzer Stra\ss e 38 01187 Dresden, Germany}
\date{\today}
\maketitle

\begin{abstract}
We use a density matrix renormalization group method to study quantitatively
the phase diagram of a one-dimensional extended Hubbard model at
half-filling by investigating the correlation functions and structure
factors. We confirm the existence of a novel narrow region with long-rang
bond-order-wave order which is highly controversial recently between the
charge-density-wave phase and Mott insulator phase. We determined accurately
the position of the tricritical point $U_t\simeq 7.2t$, $V_t\simeq 3.746t$
which is quite different from previous studies.
\end{abstract}


Much effort has been devoted to understanding the effects of strong
electronic correlations on quasi-one-dimensional systems. These materials
including conjugated polymers, Cu oxides and Ni halides exhibit rich phase
diagrams and display huge ultrafast optical nonlinearity which point to
promisinig optoelectronic applications\cite{Kiess,Ogasawara,Kishida,Iwai}.
The one-dimensional (1D) extended Hubbard model (EHM) with the nearest
neighbour repulsion $V$ in addition to the on-site repulsion $U$ is a
standard minimal model that can describe these rich physical properties.
Although this model has been investigated for more than two decades\cite
{Emery}, its ground-state phase diagram is still controversial\cite
{Hirsch,Cannon,Voit,Dongen,Zhang1,Nakamura,Tsuchiizu,Sengupta,Jeckelmann,Sandvik,Zhang2}%
. In the limit $V=0$\cite{Lieb}, the ground state is in the Mott insulating
state where the spin sector shows quasi-long-range order of spin density
wave. With use of weak-coupling renormalization group analysis\cite{Emery},
a continuous phase transition is obtained at $U=2V$ from charge-density-wave
(CDW) to spin-density-wave (SDW) phases. Strong-coupling calculations using
second order pertubation theory\cite{Emery} gave a first-order phase
transition also at $U=2V$. This means that there exists a crossover between
these two phase transitions at a tricritical point in the
intermediate-coupling regime. This picture was supported by both numerical%
\cite{Hirsch,Cannon,Zhang1} and analytical\cite{Cannon,Voit,Dongen} studies
and had been regarded for a long time as the complete phase diagram of the
1D EHM at half filling. However, the precise location of the tricritical
point where the nature of the transition changes has been a subject of much
investigations and remains uncertain\cite
{Hirsch,Cannon,Voit,Dongen,Zhang1,Nakamura,Tsuchiizu,Sengupta,Jeckelmann,Sandvik,Zhang2}%
.

Recently, by using level crossings in the excitation spectra obtained by
exact diagonalization, Nakamura\cite{Nakamura} pointed out that a novel
spontaneously dimerized phase which is also called bond-order-wave (BOW)
phase exists for small to intermediate values of $U$ and $V$ in a narrow
strip between CDW and SDW phases. This remakable proposal was confirmed by
stochastic series-expansion Monte Carlo calculations\cite{Sengupta,Sandvik}
and reformulated weak-coupling field theory\cite{Tsuchiizu} where
higher-order terms were included but was questioned by a high-level
density-matrix renomalization group (DMRG) investigation\cite{Jeckelmann}
which shows the existence of a BOW phase only on a segment of a first-order
SDW-CDW phase boundary. On the other hand, Zhang\cite{Zhang2} cast doubts on
the accuracy of the DMRG calculation and predict that a SDW-CDW phase
boundary smoothly connects the weak- and strong-coupling limits.

In order to clarify these apparent contradictions, we perform an extensive
numerical study of 1D EHM at half-filling in the vicinity of $U=2V$ using
the DMRG \cite{White} technique. In contrast to the previous DMRG studies%
\cite{Zhang1,Jeckelmann,Zhang2}, we mainly calculate the BOW correlation
functions and structure factors which are the most direct evidences for the
long-range BOW phase. Carefully using various finite-size scaling skills for
different physical quantities defined below, we confirm that the recently
discovered BOW phase\cite{Nakamura} of the 1D EHM does have a finite extent
in the $\left( U,V\right) $ plane, but its width is much smaller as has been
predicted by the Monte Carlo method\cite{Sengupta}. The position of the
tricritical point where the BOW phase vanishes is determined accurately. The
value is much higher than any previous results\cite
{Hirsch,Cannon,Voit,Zhang1,Tsuchiizu,Sengupta,Jeckelmann} while suprisingly
it is consistent with the point given in ref. \cite{Jeckelmann} where
Jeckelmann argued that the BOW phase exists up to an upper limit which is
smaller than $8t$. However the point is not identical to the tricritical
point. This can be understood as the result of increasing frustration in the
spin degrees of freedom\cite{Dongen,Jeckelmann}. Furthermore we find that
the first-order phase transition takes place for $U\lesssim 2V$ rather than $%
U\gtrsim 2V$ obtained from reformulated weak-coupling field theory\cite
{Tsuchiizu}. Using the same methods described below for $U$ up to $16t$ with
the interval of $\Delta U=0.2t$, we quantitatively determine the phase
diagram of the 1D EHM at half-filling for the first time.

The Hamiltonian of the 1D EHM is defined as follows 
\begin{equation}
H=-t\sum_lB_{l,l+1}+U\sum_ln_{l\uparrow }n_{l\downarrow }+V\sum_ln_ln_{l+1},
\label{EHM}
\end{equation}
where $n_{l\sigma }\equiv c_{l,\sigma }^{+}c_{l,\sigma }-\frac 12$, $%
n_l\equiv n_{l\uparrow }+n_{l\downarrow }$, $c_{l,\sigma }^{+}$ denotes the
creation operator of an electron at the $l$th site with spin $\sigma $ and
the bond-charge density operator $B_{l,l+1}$ is 
\begin{equation}
B_{l,l+1}\equiv \sum_\sigma \left( c_{l,\sigma }^{+}c_{l+1,\sigma
}+c_{l+1,\sigma }^{+}c_{l,\sigma }\right) .
\end{equation}

As is well-known, DMRG is a very accurate numerical method for the
ground-state properties of a 1D quantum system with short-range interactions%
\cite{White}. Here we applied the finite-size DMRG algorithm with open
boundary conditions to study the Hamiltonian (\ref{EHM}) at half-filling.
This method allows us to probe directly correlation functions and structure
factors associated with SDW, CDW and BOW in the ground state. Lattices up to 
$512$ sites were used in our studies. The largest number of states kept in
the calculation was $m=512$ per block. The hopping integral $t$ is set to $1$
as the energy unit. The weight of the discarded states was typically about $%
10^{-7}-10^{-14}$ depending on whether the system is in its critical state
or not in the final sweep. The convergence tests as functions of number of
states kept and system size were carefully performed. We checked our DMRG
calculations against exact numerical results for noninteracting ($U=V=0$)
chains (up to $512$ sites) and results from exact diagonalization for
interacting ($U\neq 0$, $V\neq 0$) chains (up to $14$ sites). Excellent
agreement was found in both cases. Furthermore, for a given chain length, we
extrapolate our DMRG results to the limit of vanishing discarded weight.
When interactions are turned on, there exist finite excitation gaps on
finite chains, so the accuracies of all quantities we calculated are not
worse than those of the noninteracting case. Thus, numerical errors in our
work could be safely estimated to be smaller than $10^{-4}$.

First, we will introduce a method which can be simply used to determine the
phase boundaries of BOW phase directly. The BOW structure factor is given by

\begin{eqnarray}
S_{BOW}(q) &=&\frac 1L\sum_{lr}e^{iqr}(\left\langle
B_{l,l+1}B_{l+r,l+r+1}\right\rangle  \nonumber \\
&&-\left\langle B_{l,l+1}\right\rangle \left\langle
B_{l+r,l+r+1}\right\rangle )~.
\end{eqnarray}
As pointed out by Nakamura\cite{Nakamura}, phase transitions on the CDW-BOW
and BOW-MI phase boundary are of Gaussian type and Kosterlitz-Thouless (KT)
type respectively, the real space staggered bond fluctuation correlation
function falls off algebraically as

\begin{equation}
\left( -1\right) ^r(\left\langle B_{l,l+1}B_{l+r,l+r+1}\right\rangle
-\left\langle B_{l,l+1}\right\rangle \left\langle B_{l+r,l+r+1}\right\rangle
)\sim r^{-\eta }
\end{equation}
Away from phase boundaries, this quantity falls off exponentially. Thus for
a finite-size system, the $S_{BOW}\left( \pi \right) $ is expected to reach
a maximum at the critical points. Fig. 1 shows the results of the $%
S_{BOW}\left( \pi \right) $ for different system sizes with $U/t=4.0$ and $%
1.5\leq V/t\leq 2.5$. As expected, the $S_{BOW}\left( \pi \right) $ peaks
twice for all the different system sizes we calculated, clearly indicating
that there exist two phase transitions. The inset of Fig. 1 shows a linear
extrapolation of these two critical values with the inverse of the chain
length $1/L$. We find that the larger the system is, the smaller the BOW
phase becomes. Nevertheless, the BOW phase remains finite at the
thermodynamic limit. Compared with other recent numerical methods\cite
{Nakamura,Sengupta,Jeckelmann}, we find this to be the most accurate way to
locate the BOW phase boundaries.

Then we will further prove the existence of a finite-width BOW\ phase
established by our method. The most direct evidence is the BOW correlation
function 
\begin{equation}
C_{BOW}\left( r\right) =\left( -1\right) ^r(\frac 1L\sum_l\left\langle
B_{l,l+1}B_{l+r,l+r+1}\right\rangle -\overline{B}^2)  \label{C-BOW}
\end{equation}
where $\overline{B}=\frac 1L\sum_l\left\langle B_{l,l+1}\right\rangle $. In
Fig. 2 (A), we show the staggered BOW correlation functions with increasing
nearest neighbour repulsion $V$ at $U/t=4.0$. To avoid boundary effects, we
only perform the average in (\ref{C-BOW}) over 256 sites in the middle of
the $512$-site system. The results indicate that there exist three different
phases in model (\ref{EHM}) since the staggered BOW correlation functions
show three distinct type of behavior as $r$ increases: (i) it decays
exponentially at $V/t=1.5$, indicating that the system has no BOW order;
(ii) it converges to a nonzero constant, indicating that the system is in
the long-range BOW phase at $V/t=2.1$; (iii) it decays as $1/r$ at $V/t=2.5$%
, indicating that the system is in the Mott insulating phase. Besides the
data we show in Fig. 2 (A), we also compare our DMRG results with Quantum
Monte Carlo calculation\cite{Sengupta} at $U/t=4.0$, $V/t=2.14$ for $L=128$
and $256$. The amplitudes of the oscillations at large distances are the
same. The BOW order parameter in the thermodynamic limit

\begin{equation}
\Delta _{BOW}=\lim_{L\rightarrow \infty }\frac 1L\sum_l\left( -1\right)
^l\left\langle B_{l,l+1}\right\rangle
\end{equation}
can be obtained by fitting $m_{BOW}\left( L\right) $ (=$\sqrt{\sum
C_{BOW}(r)/L}$) with a second-order polynomial in $1/L$ since $m_{BOW}\left(
L\right) \rightarrow \Delta _{BOW}$ for $L\rightarrow \infty $. Fig. 2 (B)
shows such extrapolations. We find that $m_{BOW}^2\left( L\right) $
approaches zero when $V/t=1.5$ and $2.5$ but remains finite when $V/t=2.1$.
The strength of dimerization is consistent with Quantum Monte Carlo
calculation\cite{Sandvik} while another recent DMRG investigation\cite
{Jeckelmann} has not detected this rather strong BOW order.

In order to give more convincing evidence, we did another finite-size
analysis in the vicinity of these two phase transitions. Let us start from
the first phase transition at $U=U_c$. Fig. 2 (C) presents plots of $\ln
[S_{BOW}\left( \pi \right) ]$ versus $\ln [L]$ for $U/t=4.0$ and three
different values of $V/t$ around the first critical point. Data points for $%
V/t=2.16$ indeed fall on a straight line, indicating critical scaling for
the BOW fluctuations. At the other two points $V/t=2.15$ and $2.17$, data
points behave nonlinearly due to the exponential decay term. Applying the
same finite-size analysis to the CDW structure factor 
\begin{equation}
S_{CDW}\left( q\right) =\frac 1L\sum_{lr}e^{iqr}\left( \left\langle
n_ln_{l+r}\right\rangle -\left\langle n_l\right\rangle \left\langle
n_{l+r}\right\rangle \right) ,
\end{equation}
we can also explore the nature of the first phase transition. The linear
behavior of $\ln [S_{CDW}\left( \pi \right) ]$ around $V/t=2.16$, shown in
Fig. 2 (D), confirms the vanishing of the charge gap at the first phase
transition point.

Next we determine the location of the phase transition of KT type. As argued
by Nakamura\cite{Nakamura}, it is a quantum phase transition of the KT type.
This makes it difficult to determine the phase boundary directly from the
behavior of $S_{SDW}\left( \pi \right) $

\begin{equation}
S_{SDW}\left( q\right) =\frac 1L\sum_{lr}e^{iqr}\left\langle
s_l^zs_{l+r}^z\right\rangle
\end{equation}
due to the finite-size effects. Instead, we apply an indirect method, used
by Sengupta et al.\cite{Sengupta,Sengupta2}, to confirm the second phase
transition. It is well known \cite{Voit} that if the ground state of a 1D
system is spin-gapless, the spin-spin correlation falls algebraically with
exponent equal to $1$. It has been further shown\cite{Clay} that in the
spin-gapless phase $S_{SDW}\left( q\right) /q\rightarrow 1/\pi $ as $%
q\rightarrow 0$ whereas in the spin-gapped phase $S_{SDW}\left( q\right)
/q\rightarrow 0$. Even a very small spin gap can be detected in this way,
since it is in practice sufficient to see the $\pi S_{SDW}\left( q\right) /q$
decay below $1$ for small $q$ to conclude that a spin gap must be present.
From Fig. 2 (E) we can see the behavior of $\pi S_{SDW}\left( q\right) /q$
for $U/t=4.0$ and different values of $V/t$. In the gapless region,
logarithmic corrections \cite{Eggert} make it difficult to observe the
approach to $1$ as $q\rightarrow 0$. In analogy with spin systems\cite
{Eggert2}, we expect the leading logarithmic corrections to vanish at the
point where spin gap opens and therefore exactly at the critical point there
should be a clear scaling to $1$. Based on results shown in Fig. 2 (E), we
estimate the BOW-MI boundary to be at $V/t=2.0\pm 0.01$ at $U/t=4.0$ which
is consistent with the results shown in Fig. 1. The Fig.2 (F) further
provides evidence for the transition from the spin-gapped state, as
identified by the exponential decay of the staggered SDW correlation
function, to the spin-gapless state, as characterized by the $1/r$-decay of
the SDW correlation function

\begin{equation}
C_{SDW}\left( r\right) =\frac 1L\left( -1\right) ^r\sum_l\left\langle
s_l^zs_{l+r}^z\right\rangle .
\end{equation}

Now Let us determine the tricritical point. As we mentioned above, Nakamura%
\cite{Nakamura} first argued that there exists an extended BOW phase for
couplings weaker than a tricritical point, which is supported by numerical%
\cite{Sengupta} and analytical\cite{Tsuchiizu} studies where the tricritical
point is given as $U_t\approx 4.76t$ and $U_t\approx 5t$ respectively.
However, recent DMRG calculation\cite{Jeckelmann} disagreed this picture and
pointed out that the BOW phase exists only on the first-order SDW-CDW
phase-transition line for intermediate coupling $U$ starting from the
tricritical point $U_t\approx 3.7t$ to an upper limit which is smaller than $%
8t$. In order to give a more convincing result, we have studied the phase
transitions in a wide range of coupling $U$ starting from $2t$ to $9t$ with
the interval of $\Delta U=0.2t$ as carefully as we displayed above. We find
that the width of the BOW phase will shrink to $0$ at $U\approx 7.2t$ in the
thermodynamic limit and the extrapolation of the BOW order parameter will go
to $0$ for all the couplings $V$ at $U>7.2t$ which indicate the vanishing of
long-range BOW phase and the merging of these two continuous phase
transitions into one. Furthermore, we also calculate the $V$ dependence of
the CDW order parameters $m_{CDW}\left( L\right) $ (the definition is
similar to $m_{BOW}\left( L\right) $) across the phase boundary for $U$ up
to $16$. From the inset of Fig. 3 we can see that the characteristics of a
second-order transition for $U=6t$ and a first-order transition for $U=9t$
are indeed quite apparent. The points are obtained from calculating the CDW
order parameters for the chain length $L=192$, $256$, $320$, $512$ and then
extrapolating to the thermodynamic limit $L\rightarrow \infty $. The
transition point $V_c\approx 4.143$ for $U=8t$ and $V_c\approx 6.116$ for $%
U=12t$ are consistent with previous results\cite{Sengupta,Jeckelmann}.

The phase diagram obtained from DMRG calculations described above is shown
in Fig. 3. The BOW phase exist from weak coupling up to tricritical point
where the two continuous transition line meet. With the increasing of $U$
and $V$ along the line $U=2V$, the BOW phase first expands and then shrinks
up to the tricritical point. Beyond this point the BOW phase disappears and
the first order transition from the CDW to SDW phase can be obtained. This
phase diagram is similar to the schematic one in ref. \cite{Sengupta} except
that the width of BOW phase become much smaller and extend to much larger
coupling. The first-order phase transition takes place for $U\lesssim 2V$
rather than $U\gtrsim 2V$ reported in ref. \cite{Tsuchiizu}. Our result does
not agree with the location and the width of the BOW phase given in ref. 
\cite{Jeckelmann} while the upper limit of the BOW phase is consistent with
each other.

In conclusion, we have studied 1D half-filled EHM using the DMRG method. The
quantitative phase diagram is obtained by investigating correlation
functions and structure factors for the first time. The novel spontaneously
dimerized phase does have finite width in the $\left( U,V\right) $ plane.
Our estimate for the tricritical point where the BOW phase vanishes and the
nature of the transition changes is $U_t/t\approx 7.2$, $V_t/t\approx 3.746$%
. The extension of dimerization to rather strong coupling can be understood
as the results of increasing frustration in the spin degrees of freedom\cite
{Dongen,Jeckelmann}.

It is our pleasure to acknowledge helpful discussions with Dr. P. Thalmeier.

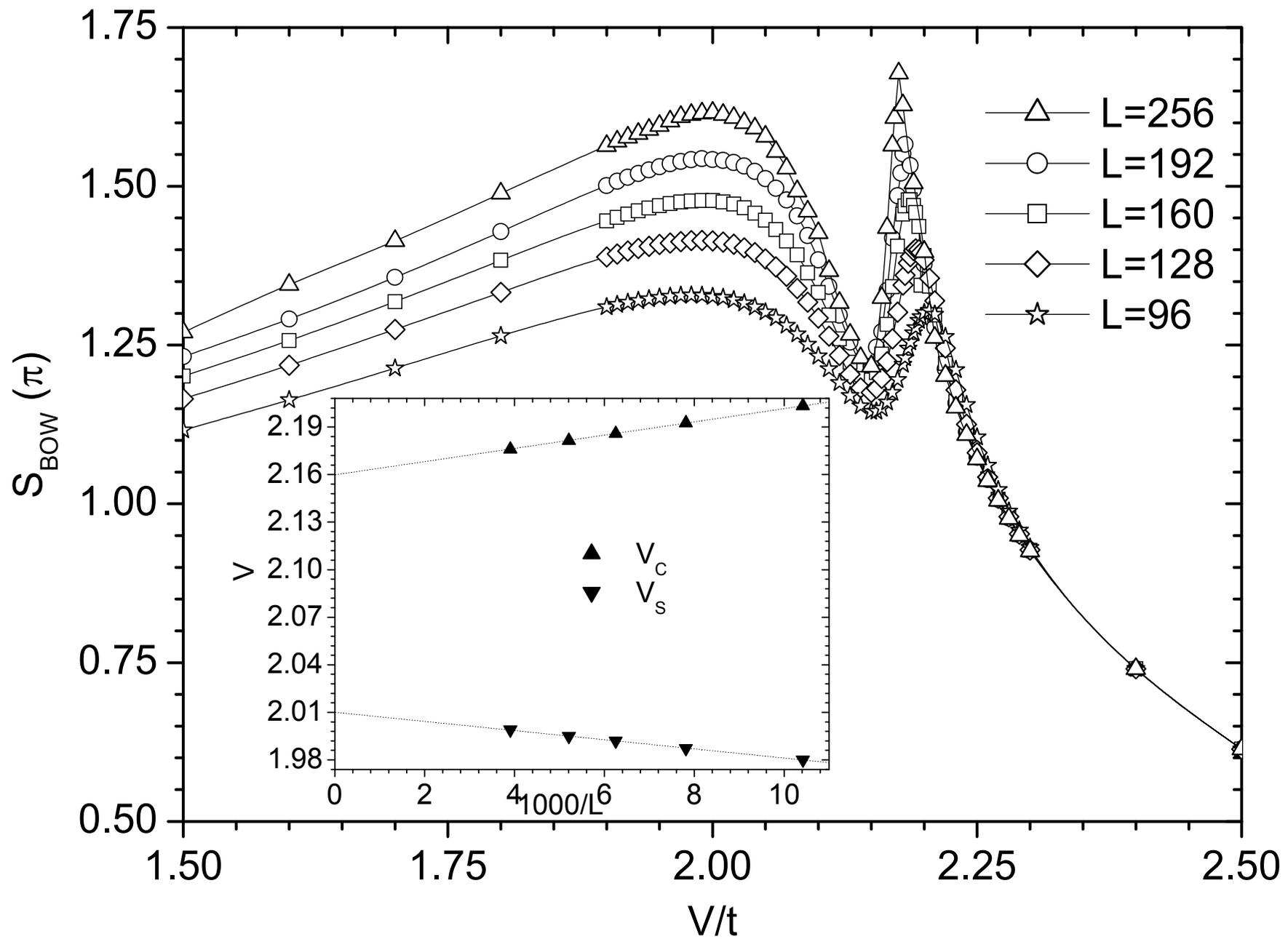
\begin{figure}[tbp]
\caption{Behavior of $S_{BOW}(\pi)$ across the phase boundary for $U/t=4.0$.
The inset shows a linear extrapolation of the critical values $V_c$ and $V_s$
with the inverse of chain length $1/L$.}
\end{figure}

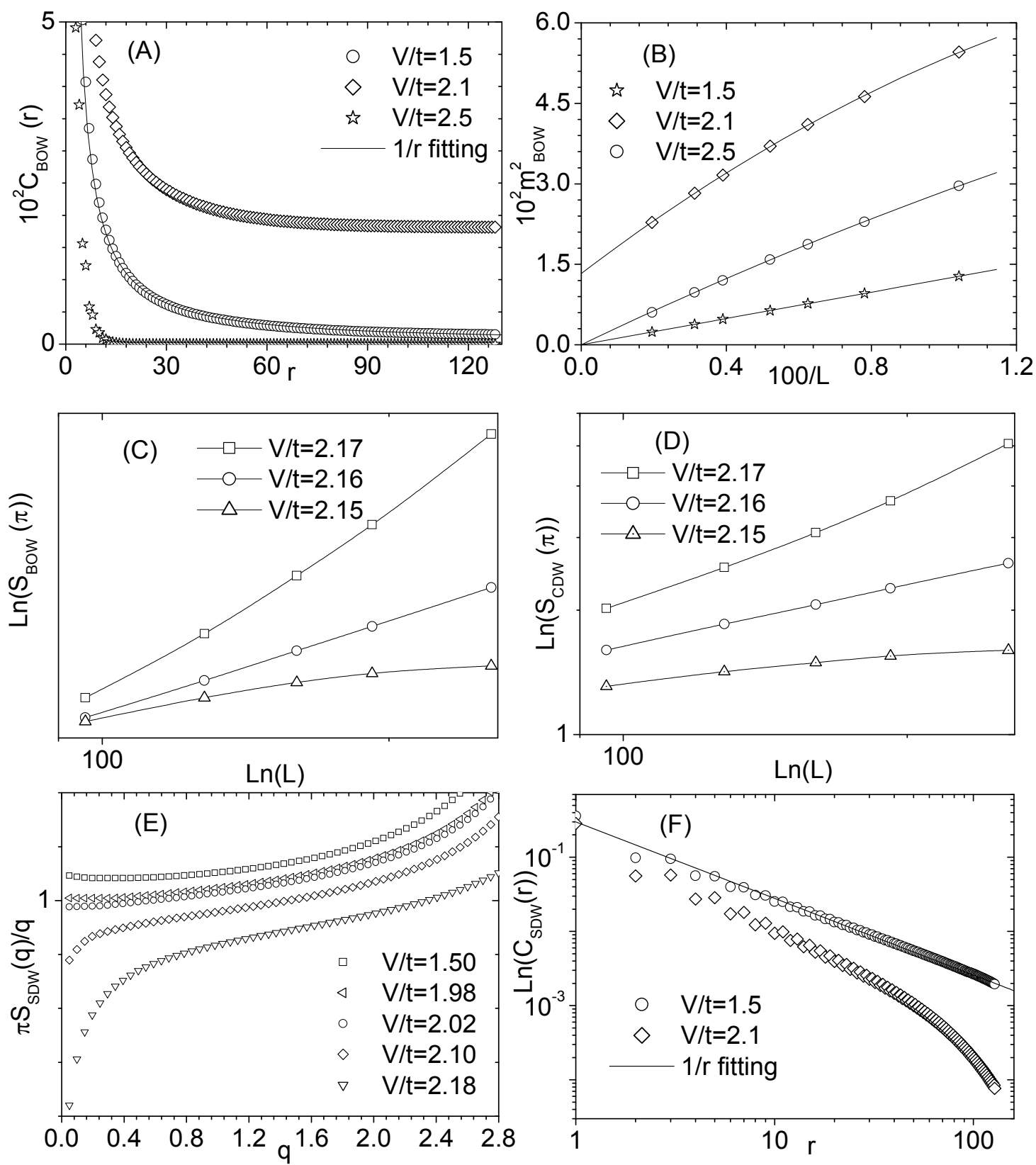
\begin{figure}[tbp]
\caption{Finite-size analyses for BOW, CDW, and SDW correlation functions
and structure factors in the vicinity of the two phase transitions at $V_c$
and $V_c$ for $U/t=4.0$.}
\end{figure}

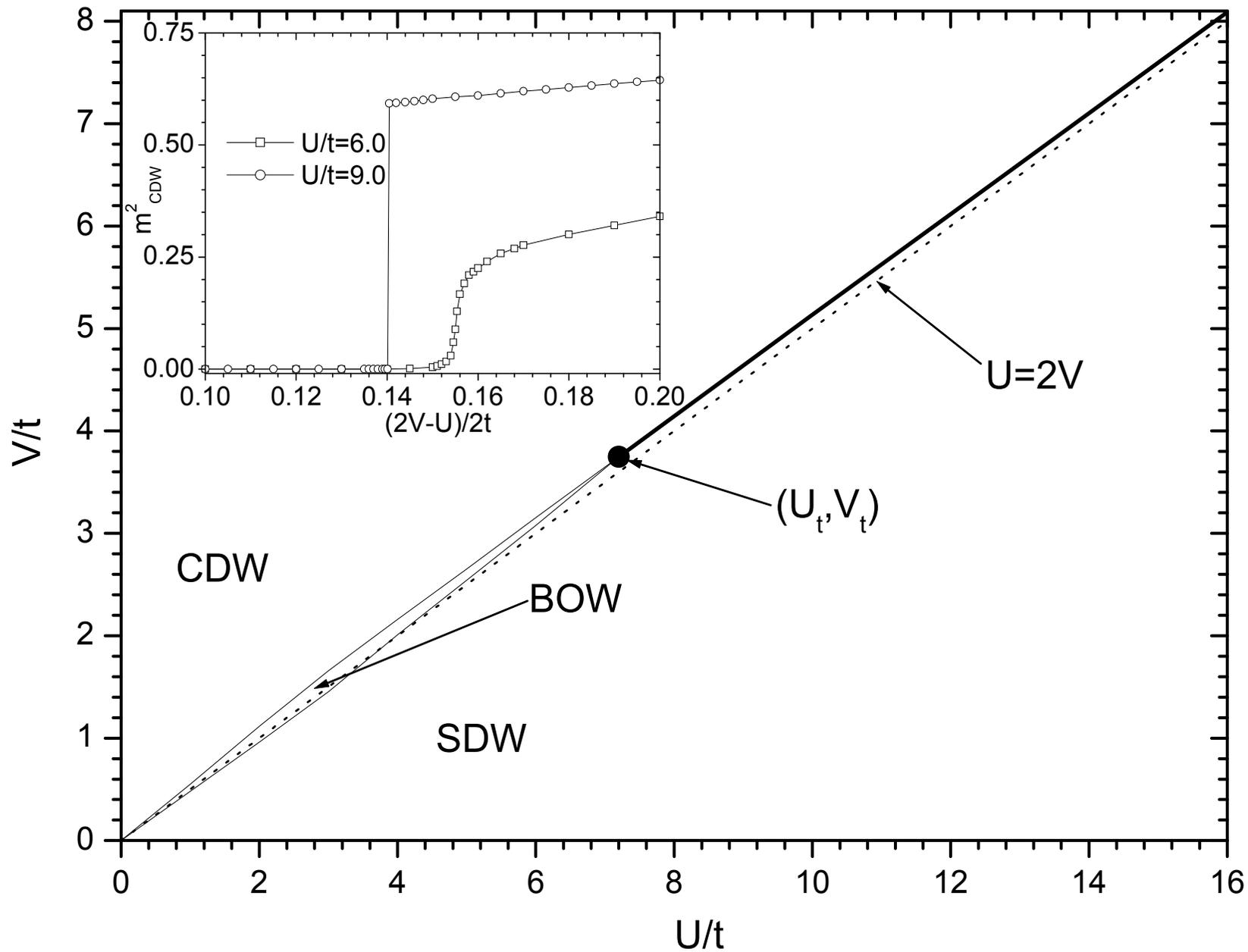
\begin{figure}[tbp]
\caption{Phase diagram of the half-filled extended Hubbard model. The
tricritical point is at $(U_t,V_t)\simeq (7.2t,3.746t)$. The inset shows $V$
dependence of the CDW order parameters $m_{CDW}$ across the phase boundary
for $U=6t$ and $U=9t$ in the thermodynamic limit.}
\end{figure}

\end{document}